# Ion-wake Field inside a Glass Box

Mudi Chen[1], Michael Dropmann[1,2], Bo Zhang[1], Lorin S[1]. Matthews and Truell W. Hyde[1]

[1]*Baylor University, Center for Astrophysics, Space Physics and Engineering Research*
*One Bear Place 97310, Waco, Texas 76798-7310, USA*
[2]*Institute of Space Systems, University of Stuttgart*
*Raumfahrtzentrum Baden-Württemberg, Pfaffenwaldring 29 ,70569 Stuttgart, Germany*
e-mail address: Mudi_chen@baylor.edu, Truell_Hyde@baylor.edu

The confinement provided by a glass box is proving ideal for the formation of vertically aligned structures and a convenient method for controlling the number of dust particles comprising these dust structures, as well as their size and shape. In this paper, the electronic confinement of the glass box is mapped and the particle interactions between the particle pairs inside the glass box are measured. The ion-wake field is shown to exist within the glass box and its vertical and horizontal extent is measured.

# I Introduction:

A dusty plasma [1] is best described as a weakly ionized gas containing electrons, ions, neutral atoms and small solid particles, usually consisting of micron-size spheres. Due to the high thermal speed of the electrons, these micro particles are usually negatively charged [2, 3]. The charge on any single particle can vary from a few hundred to several thousand elementary charges, depending on particle size and plasma conditions. In an experimental setting on the earth, the particles can be levitated against the force of gravity in the plasma sheath near the powered electrode by a self-induced electric field. Depending on the external confinement, one-dimensional (1D) to three-dimensional (3D) dust structures can be formed [4-7]. Placing a glass box on the lower electrode is a common method for providing such confinement and is used to create dust clusters with different numbers of particles, from one to several dozens, by tuning the RF power. Once established, transitions from 1D to 2D zigzag, 2D to 3D helical structures and helical structures to layered structures [8] are easily obtained.

Particles also interact directly with one another through screened Coulomb repulsion and indirectly by altering the ion flow, which determines the plasma screening and ion-wake field. The ion-wake effect has been studied both theoretically [9-16] and experimentally [17-20]. Research has shown that charged dust particles tend to align with the ion flow [10, 13]. Upstream particles focus the ions at a point beneath them, and downstream particles are attracted by the positive space charge region created due to this ion focusing. A nonreciprocal interaction [9, 19] between the upstream and downstream particles is a signature property of the ion-wake effect. In this case, upstream particles dominate the motion of downstream particles, while the reverse effect is so small that it can often be considered negligible. Recent simulation studies [21] also indicate that the downstream particles can become discharged by this interaction between the upstream and downstream particles.

As noted, a glass box confinement has proven ideal for the formation of vertically aligned 1D, 2D and 3D structures, which are difficult to obtain under other types of confinement. However, no study of the role the ion-wake effect plays within such a confinement or whether the ion-wake field exists at all in this environment has been conducted to date. Given the strong confinement produced by the glass box and the direct interaction between the particles, an investigation of the ion-wake effect has long been ignored, leaving a proper explanation of the underlying physics sorely needed.

In this paper, under the confinement created by a glass box, the interaction between the particles and the ion-wake field as it pertains to vertically aligned single particle chains is examined. Section II provides a short description of the experimental setup employed, while Section III shows data collected mapping both the accelerations provided by the confining forces within the glass box as well as the mutual particle accelerations as a function of the particles' locations relative to each other. A discussion of this data, which shows the existence of the wake field, is given in Section IV, with conclusions presented in Section V.

## II Experimental Setup

The experiment described here was performed in a modified gaseous electronics conference (GEC) radio-frequency (RF) cell, filled with argon at a pressure of 6.67 ±0.10 Pa. An RF electrical field was produced by a pair of capacitively-coupled electrodes 8 cm in diameter, situated one above the other, and separated by a distance of 2.54 cm. The upper electrode was grounded, while the lower electrode was powered by a RF generator at a constant frequency of 13.56 MHz. The amplitude of the input RF signal was 2 W. An open-ended glass box of dimension 1.27 cm $\times$ 1.07 cm $\times$ 1.07 cm (height $\times$ width $\times$ length) with 2 mm wall thickness was placed at the center of the lower electrode. Melamine formaldehyde spheres having a manufacturer-specified mass density of 1.514 g/cm$^3$ and diameter of 12.00 ± 0.09 μm were used. A dust dropper was employed to introduce the particles into the glass box, where they were illuminated by a vertical sheet of laser light. The particles' positions were recorded at 500 or 1200 frames per second (fps) using a side-mounted, high-speed CCD (Photron) camera and a microscope lens. In order to manipulate isolated particles, optical radiation from a Coherent Verdi V-5 laser was introduced into the chamber using an adjustable optical system. The power supplied to the laser was user controlled and held between 0.01 ~ 1.50 W.

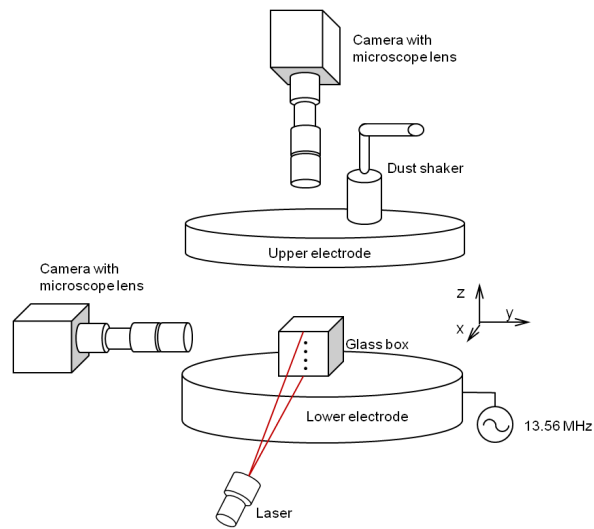

Fig.1 Experimental setup. The top electrode is grounded while the bottom electrode is powered. The separation distance between the electrodes is 2.54 cm. The open-ended glass box shown has dimensions of 1.27 cm × 1.07 cm × 1.07cm. A Coherent VERDI V-5 laser is employed to perturb individual particles, the motion of which is then captured using a Photron CCD high speed camera (side view).

## III Data Collected and Results

1. **Confinement**

Since the dust particles are levitated either in the sheath or pre-sheath region within the small volume of the glass box, a technique producing only small perturbations is required to measure the plasma environment around the dust particles. In this case, a free-falling-particle technique was applied to measure the confinement produced by the glass box. In this technique, the dust particle itself is used as a probe to measure the forces affecting the particles and create a map of the overall confinement. In our experiment, a few hundreds of dust particles were dropped from the dust dropper located above the upper electrode into the glass box, with individual particle motion recorded at 1200 fps. A MATLAB-based algorithm for particle detection was employed to identify the particles in each frame of the data collected, which were then linked using the Hungarian algorithm [22].

The resulting particle trajectories were examined to eliminate any faulty data. Trajectories exhibiting discontinuities in their calculated accelerations were removed as were trajectories exhibiting minimal particle movement, since both of these are usually associated with noise falsely interpreted as particles. Particle-particle interactions were managed by establishing a minimum allowable distance between detected particles. For this case, it was 1 mm.

Based on the first and second order difference quotients for the trajectories of the particles, the velocities and accelerations in both the vertical ($\dot{Z}, \ddot{Z}$) and horizontal ($\dot{X}, \ddot{X}$) directions were determined. Assuming that only gas drag, confining electrostatic forces and gravity are acting on the particles, the equations of motion are given by

$$m_d \ddot{Z} = \beta \dot{Z} + Q_d E_z - m_d g \qquad (1)$$
$$m_d \ddot{X} = \beta \dot{X} + Q_d E_x \qquad (2)$$

where $m_d$ is the mass of the dust particle, g is the acceleration due to gravity, $Q_d$ is the charge on the dust particle, and $E_z$ and $E_x$ are the electric fields in the vertical and horizontal directions. Since the ion drag force is small compared to the confinement force and no external thermal gradient is applied to the system, the ion drag and thermophoretic forces may be considered negligible [23,24] and then not included. In Eqs. (1) and (2), $\beta$ is defined as

$$\beta = \delta \frac{4\pi}{3} a^2 N m_n \bar{c}_n \qquad (3)$$

where a is the radius of the dust particle, N is the neutral gas number density, $m_n$ is the mass of the neutral gas atoms (Argon), and the coefficient $\delta$ accounts for the microscopic mechanism governing collisions between the gas atom and the surface of the dust particle. For this experiment, with melamine formaldehyde (MF) particles and Argon gas, $\delta$ has been determined to be 1.44 as reported in Ref [25,26]. Finally, $\bar{c}_n$ is the thermal speed and defined as

$$\bar{c}_n = \sqrt{\frac{8kT}{\pi m_n}} \qquad (4)$$

where k is the Boltzmann constant and T is the temperature.

For the given experiment at conditions employed, Eqs. (3) and (4) yield a value for $\beta$ = 9.26×$10^{-12}$ N $m^{-1}$ s. This coefficient was also measured experimentally [27], giving $\beta$ = (9.77 ±2.56)×$10^{-12}$ N $m^{-1}$ s, in good agreement with the analytical value. The analytical value was used in all following calculations.

Employing this result and using Eq. (1) and (2), the electrostatic forces, $Q_d E_z$ and $Q_d E_x$, in the vertical and horizontal directions can now be calculated. For convenience, this data was recorded as the vertical and horizontal acceleration $Q_d E_z/m_d$ and $Q_d E_x/m_d$.

These calculated particle accelerations were mapped as a function of position within the box. Since this paper focuses on 1-D vertically aligned dust particle structures confined at the center of a glass box, only acceleration maps for the central region of the glass box were generated as shown in Figs. 2(a) and (b). This figure covers the horizontal and vertical range of all recorded particle trajectories with the vertical displacement measured from the top of the glass box and the horizontal displacement measured from the center of the box. The mapping used a grid spacing of 0.1 mm, and the acceleration at each grid point was calculated as the average acceleration for all data points within a radius R = 0.05 mm of each grid point. To minimize effects due to particle interactions, only data points from particles at least 1mm away from any other particle were included.

Fig. 2(a) shows the horizontal acceleration of the dust particles. This acceleration is always directed toward the center of the box, with the greatest acceleration of approximately 0.5 g observed at the top edge. As shown, in the central region of the box (-1 mm ≤ x ≤ 1 mm, -2 mm ≤ z ≤ -6.5 mm) the horizontal confinement force can be treated as a linear force with a restoring constant of -0.11 ±0.01 $m_d$ g $mm^{-1}$.

Fig. 2(b) shows the vertical acceleration map for the dust particles. Particles can only levitate in the region where their vertical acceleration is greater than or equal to the acceleration due to gravity. In contrast to the conditions within the plasma sheath formed without a box, under the experimental parameters shown, the vertical confinement force inside the glass box does not increase monotonically as a particle approaches the lower powered electrode. A maximum vertical acceleration in magnitude of ~1.05g is found at the middle of the box (-2 mm ≤ x ≤ 2 mm), with an extended vertical region (5.5 mm ≤ z ≤ 9 mm) where the acceleration is approximately -1g. As such, a 1D dust string consisting of more than two particles can only be formed in such a region. For the operating parameters used to generate the data shown in Fig.2, the longest particle chain observed consisted of 11 particles spanning a vertical region of 9.7 mm. It should be noted that unlike the case without a box, where the radii of the particles within such a vertical structure are often different, a particle chain formed inside the box consists of nearly identically sized particles.

The vertical levitation region is highly dependent on the system's operating parameters. For example, decreasing the RF power decreases both the extent of the region as well as the magnitude of the confining force until at some critical points, particles can no longer be levitated and those closest to the lower electrode are removed. Thus, the number of particles comprising a vertically aligned particle chain can be controlled using the RF power.

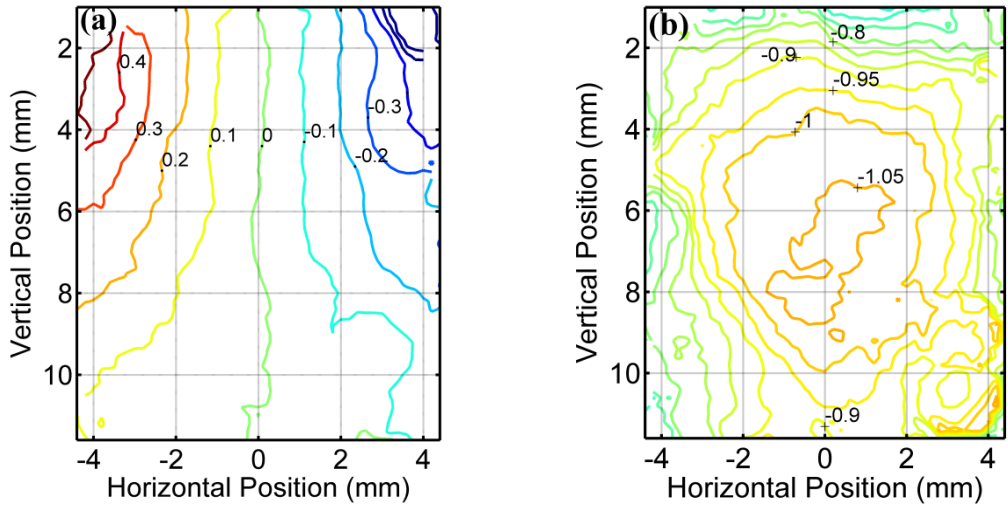

Fig. 2 Maps of the horizontal (a) and vertical (b) accelerations measured within the box. Experiments shown were run at 6.67 Pa Argon gas pressure and 2 W RF power. Particle acceleration is given as a multiple of g and the negative sign indicates upward acceleration. In the vertical direction, the distance is measured from the top of the glass box. In the horizontal direction distance is measured from the center of the box.

## 2. Particle-Particle Interaction within a Two-Particle-Chain

In order to examine the particle-particle interactions within the chain, a Coherent VERDI G5 laser was employed to perturb individual particles and their subsequent motion was recorded using a high-speed camera running at 500 fps. In the simplest case, a particle pair was formed and an individual particle was pushed horizontally using a pulsed laser beam of 100 ms duration with 10 mW power. The diameter of the beam spot was approximately 50 μm and the particles were initially aligned vertically with an interparticle separation distance of 0.80 ±0.05 mm. Due to the short heating time, low laser power, and small diameter of the beam as compared to the interparticle distance, each particle could easily be perturbed separately. Fig. 3 shows the experimental data as well as a cartoon of the particle motion while separately perturbing the top or bottom particle.

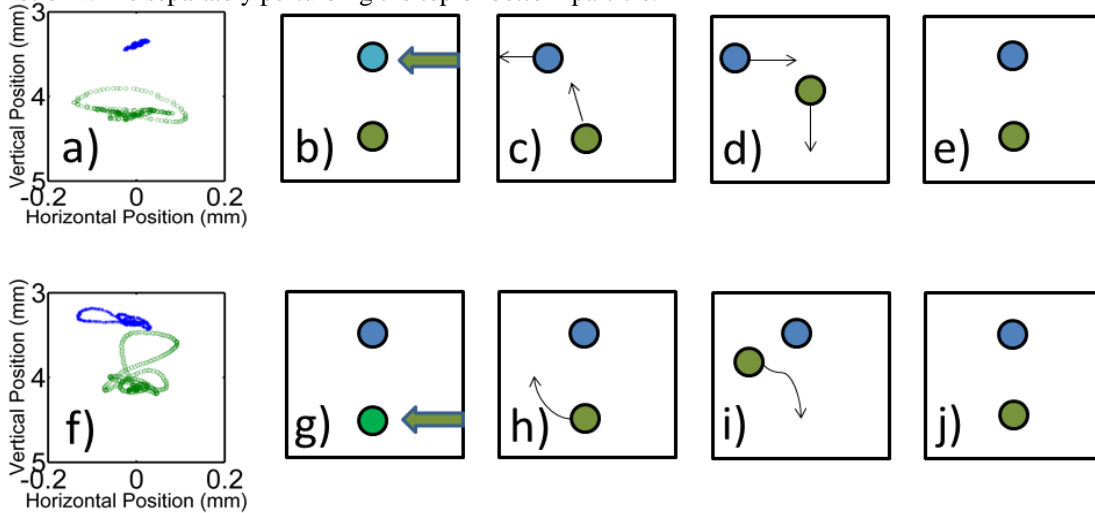

Fig. 3 Perturbed motion of particles within a particle pair. Blue represents the top particle of the pair while green represents the bottom particle. The experimental data (a) and representative cartoon (b-e) when only the top particle is disturbed; the experimental data (f) and representative cartoon (g-j) when only the bottom particle is disturbed. The vertical axis and horizontal

axis in (a) and (f) represent the distance to the top and the middle of the box, respectively.

As shown in Fig. 3 (b)-(e), pushing the top particle to the left causes the bottom particle to move upwards and slightly to the left. As the horizontal displacement of the top particle increases, the confinement produced by the box eventually forces it back toward equilibrium, resulting in the bottom particle returning to its original position. Perturbation of the bottom particle in an analogous manner is illustrated in Fig. 3 (g)-(j). Once the bottom particle is removed from its equilibrium position, it immediately moves upwards as shown in Fig. 3(h). During this time, the top particle remains stationary until the interparticle distance is smaller than 550 μm, at which point it is repelled slightly upwards and to the right. Once the bottom particle returns to its equilibrium position (again due to the horizontal confinement), it drops quickly beneath the top particle, as seen in Fig. 3(i). Finally, in Fig. 3(j), the particle pair resumes its stable configuration after a short time of exhibiting damped oscillations.

The total force acting on the particles once the laser beam is removed consists of the confinement force, the neutral drag force and the particle-particle interaction force

$$\boldsymbol{F}_{total} = \boldsymbol{F}_{conf} + \boldsymbol{F}_{drag} + \boldsymbol{F}_{inter} \tag{5}$$

since for the reasons described earlier, the ion drag force and the thermophoretic force can be considered negligible. Therefore, using the previously measured confinement force and total force while assuming a neutral drag force given by

$$\boldsymbol{F}_{drag} = -\beta \boldsymbol{v}, \tag{6}$$

the particle-particle interaction force can be calculated from the measured total force from Eq. 5.

Fig. 4 shows the displacement (Fig. 4(a)(c)) and the calculated acceleration (Fig. 4(b)(d)) of both particles due to the particle-particle interaction force in the horizontal direction. Negative quantities represent movement directed toward the left, while positive numbers represent motion to the right. Note that the negative acceleration of the illuminated particle persists for some time after the laser is turned off. This is assumed to be due to the photophoretic force caused by particle heating [28]. As shown, when the top particle is perturbed from its equilibrium position (Fig. 4(a)), the bottom particle follows in the same direction over a short range (100μm) before reversing direction to oscillate about its equilibrium position.

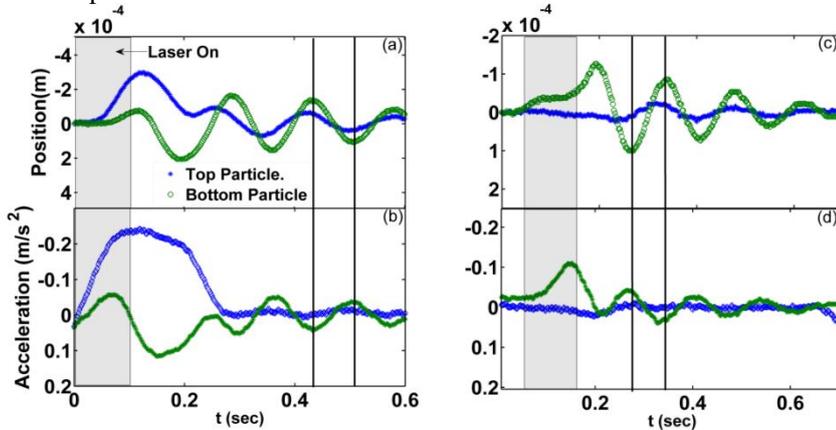

**Fig. 4 Particle motion and acceleration due to the particle-particle interaction force in the horizontal direction. (a) Motion and (b) acceleration for perturbation of top particle and (c) Motion and (d) acceleration for perturbation of the bottom particle for the conditions described in the text. The shaded region indicates the time during which the laser illuminated the particle. Vertical**

lines indicate the positions and accelerations of particles at corresponding times.

When the bottom particle is pushed to left, as shown in Fig. 4(c), the top particle remains at its original position before being repelled slightly to the right. This phenomenon confirms the assumption that the initial motion observed for the bottom particle when the top particle was pushed (see the trajectory during the first 0.1 s in Fig. 4(a)) was not caused by the laser. The damped particle oscillation also allows the neutral gas drag coefficient $\beta$ to be calculated using [29],

$$\ddot{X} + \beta\dot{X} + \omega_0^2 X = 0 \qquad (6)$$

where X is the horizontal displacement from equilibrium position and $\omega_0$ is the natural frequency of the horizontal potential well. The value of $\beta$ determined in this manner is $(8.97 \pm 1.15) \times 10^{-12}$ N m$^{-1}$ s, also in agreement with analytically calculated results.

Figs. 4(b) and (d) show the acceleration created by the force imparted by the laser and resulting interparticle interactions. In the first case, the top particle was initially given an acceleration of 0.3 m/s$^2$ to the left by the laser. The bottom particle followed the top particles, was displaced to the left and experienced a maximum acceleration of 0.05 m/s$^2$. As the particles returned to oscillate about their equilibrium positions after 0.3 s, it can be seen that the acceleration of the top particle was almost zero, while the bottom particle still had a maximum acceleration of 0.04 m/s$^2$ directed toward the top particle, as indicated by the vertical lines showing the correlation between the positive force for a negative displacement. Upon perturbation of the lower particle, Fig. 4(d), the top particle was repelled, moving to right for the first 0.2 s during the time that the bottom particle approached it from below and left. By the time the bottom particle returned to its equilibrium position, the interaction force acting on the top particle was almost negligible, although the bottom particle continued to exhibit an attractive interaction, again indicated by the vertical lines.

Using the perturbation method described, acceleration maps for the interparticle interaction between the top and bottom particles were generated by separately perturbing particles with varying laser powers between 0.01W to 0.50W. The region where the data was collected was overlaid with a grid with spacing of 0.1mm and the average acceleration was calculated for all data points within a radius of 0.05mm of each grid point. Accelerations were calculated only for regions containing at least five data points. In all cases, the coordinate origin was centered on the position of the top particle, indicated by blue dot, while the y-axis and the x-axis represent the vertical and horizontal interparticle separations respectively. At equilibrium, the bottom particle is located at [0, 0.79], indicated by the green dot, measured with respect to the location of the top particle.

In Fig.5 (a) and (c), positive values indicate horizontal acceleration to the right, while in Fig. 5 (b) and (d), positive values represent downward vertical acceleration. Fig. 5(a) and (b) provide interaction maps for the force of the top particle acting on the bottom particle, (generated by perturbing the top particle) while the data shown in Fig. 5(c) and (d) were obtained by perturbing the bottom particle and show the force exerted on the top particle by the bottom particle at the specific separations. Taken together, these provide the interparticle interactions between the bottom and top particles. Note that the horizontal axes are asymmetric in the two cases, as the bottom particle is to the right of the top particle when the top particle is perturbed, while the bottom particle is to the left of the top particle when the bottom particle is perturbed. The white regions out of the contours indicate lack of data.

In Fig. 5(a), the horizontal acceleration of the bottom particle is shown obtained for the region where the interparticle distance was larger than 0.4 mm. The horizontal accelerations are always directed toward the midline directly beneath the top particle, indicating that the bottom particle experiences a horizontal attractive interparticle interaction across this region. The strongest attractive acceleration, -0.05 m/s$^2$, is observed approximately 0.3 mm away from the midline beneath the top particle, which is in agreement with measurements made by Hebner using two particles with different mass approaching each other under a cutout confinement[18, 30]. As seen in Fig. 5(b), the vertical acceleration of the bottom particle is always downward, showing there is only a repulsive interaction from the top to the bottom particle. The maximum downward acceleration, 0.31 m/s$^2$, is observed at approximately [0, 0.79], which coincides with the particle's equilibrium position. As shown in Fig. 5(c), the horizontal acceleration provided from the bottom to top particle is almost zero when the bottom particle is at its equilibrium positions. All the accelerations in the Fig. 5(c) are positive (to the right) while the bottom particle is at left side of the top particle. This indicates that the bottom particle repel the top particle to right. Note that the top particle can only be repelled by the bottom one, and that repulsive acceleration increases with decreasing interparticle distance. In Fig. 5(d), the bottom particle causes the top particle to have an upward (negative) acceleration when their vertical separation is larger than 0.2mm. In this region, the repulsive interaction from bottom to top is increasing with decreasing distance with a maximum magnitude less than 0.1 m/s$^2$. It is interesting to note that the top particle gains a downward acceleration when the bottom particle is close to it and in the same horizontal plane

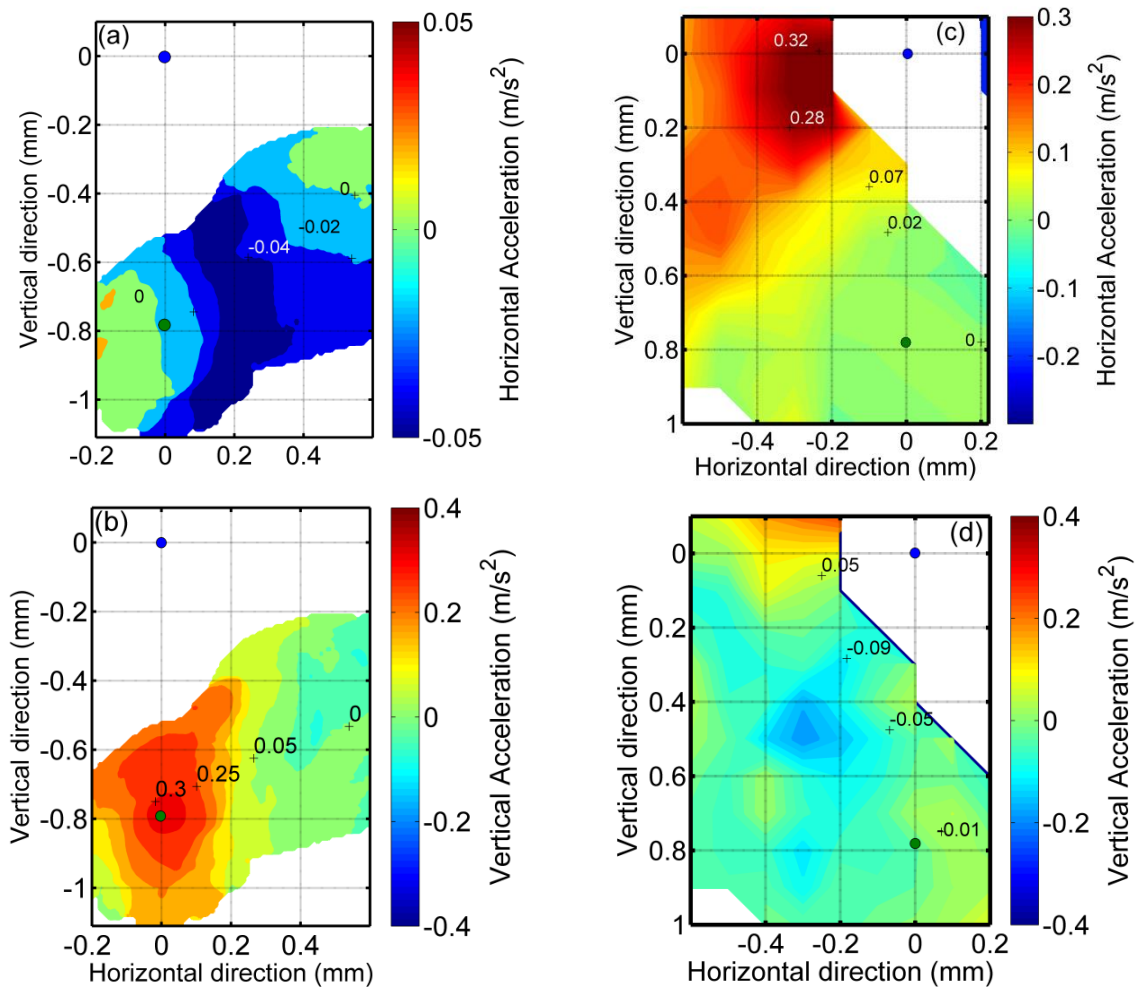

**Fig. 5** Maps of the acceleration due to interparticle interaction. Figures (a) and (b) show the horizontal and vertical acceleration of the bottom particle due to the interaction from the top particle when the top particle is perturbed. Figures (c) and (d) show the horizontal and vertical acceleration of the top particle due to the interaction from bottom particle when the bottom particle is perturbed. In all cases, [0,0] is the position of the top particle and the x- and y-axes give the horizontal and vertical displacement of the bottom particle from the top particle. Positive acceleration values indicate rightward and downward acceleration, respectively. The blue point indicates the position of the top particle at the point [0, 0], while the green circle indicates the equilibrium position of the bottom particle [0, 0.079].

## IV Discussion

As shown in Figs. 4, 5(a) and (c), in the horizontal direction, the top particle of a vertically aligned two particle chain attracts the bottom particle, while the bottom particle exerts little to no force on the top particle at the same relative positions. Only repulsive horizontal force can be observed on the top particle from the bottom particle in Fig. 5(c). At the same time, Figs. 5(b) and (d) show that while both the top and bottom particles in the chain repel each other in the vertical direction, the acceleration of the top particle is much smaller than the acceleration of the bottom particle. The top particle exerts a maximum downward vertical acceleration (and therefore force) of 0.31 m/s$^2$ on the bottom particle when it is at a point 0.79 mm directly below the top particle. This point coincides with the equilibrium position of the lower particle. The corresponding acceleration of the top particle in the equilibrium configuration is less than 0.01 m/s$^2$. In other words, the interaction between the top and bottom particle is non-reciprocal in both the horizontal and vertical directions.

This non-reciprocal attractive interaction in the horizontal direction has been previously observed in particle-pair experiments under similar operating conditions but without a glass box confinement [19,20]. It is generally explained using the ion-wake effect, where a positive ion-focusing region is formed beneath the upper particle due to the ion flow, as shown in Fig. 6. The positive space charge region, the location of which is determined by the upstream particle, can attract the downstream particle, while the downstream particle can only repel the upstream particle since both are negatively charged. The positive space charge region also causes the downstream particle to charge less negatively as there is an enhanced ion impact on the downstream particle. This discharging effect has been measured experimentally [20] as well as predicted numerically [21, 31] but again without a glass box confinement. Given that a non-reciprocal interaction is considered to be a primary indicator of the ion-wake field effect, the experiments here prove the existence of an ion-wake field within the box.

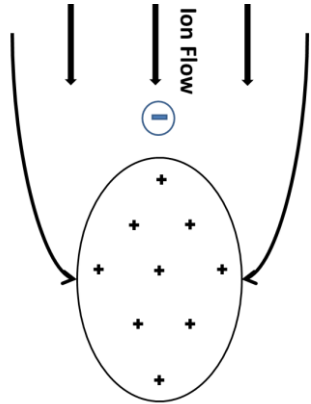

**Fig. 6** Scheme of the ion wake. The blue circles is the dust particle

It is interesting to note that due to the strong confinement provided by the box, an attractive interaction can only be observed close to the midline beneath the top particle, which in these experiments spans a horizontal region -0.3 mm ≤ $x$ ≤ 0.3 mm. Outside this region, the horizontal confinement produced by the glass box is much larger than the observed attractive effect, which may explain why the existence of the ion-wake effect within a glass box has remained in question for such a long time. In the vertical direction, the wake field effect is much more apparent. As shown in Fig. 5(b) and (d), the bottom particle always experiences a downward interaction while providing only a small interaction acceleration on the top particle, particularly at vertical distances larger than 0.7 mm. This implies that the non-reciprocal interparticle ratio (i.e., the interaction between the upstream and downstream divided by the interaction between the downstream and upstream particle) may be quite large (approximately 5-60). This is interesting since for experiments without a glass box, this ratio is much smaller (approximately 5-10) [20]. The largest value found for this ratio in there is R = 60, observed when particles are near their equilibrium positions (top particle at [0, 0] and bottom particle at [0, 0.79]). The downward acceleration gained by top particle when the bottom particle is close to the top particle maybe due to the resolution of our mapping technique.

# V Conclusions

A particle free-fall technique has been employed to map the confinement produced by a glass box placed on the lower powered electrode of a GEC rf reference cell. Interaction maps between the top and bottom particle and the bottom and top particle of a two-particle chain were generated employing a laser beam to perturb each individual particle. A non-reciprocal particle interaction was observed in both the horizontal and vertical directions, confirming the existence of an ion-wake field within the glass box. It was shown that for this case, the predicted attractive horizontal non-reciprocal phenomenon can only be observed near the midline of the box due to the existence of the strong inherent confinement forces produced by the box while in the vertical direction, the non-reciprocal ratio is much greater than that observed without a glass box confinement. This explains why the existence of an ion-wake field in a glass box has long been in question; the horizontal attractive interaction between a particle and the ion-wake field is much smaller than that provided by the confinement force for most regions within the box. However, at the midline of the box, these become comparable and in the vertical direction, the effect of the ion-wake field can become large enough to play an integral role in determining the position of the downstream particle. A representative example of this is the extended levitation region observed where dust particles experience an upward electrical force approximately equal to that of the gravitational force. This 'flat' confinement region is the key factor for forming long 1D vertically aligned dust particle structures.

The above shows that additional confinement data will be required in order to enhance the overall resolution and provide a more detailed interaction map. Once completed, this will allow investigation of the interaction between the ion-wake and particle structure formation. This research is currenetly underway and will be reported in a future publication.

# Acknowledgement

Support from NSF grant #1262031 and NSF/ DOE grant #1414523 is gratefully acknowledged.